\documentclass[12pt]{article}

\usepackage{epsfig}
\usepackage{amsmath}

\begin{document}

\title{Light-scattering spectrum of a viscoelastic fluid subjected 
to an external
temperature gradient}
\author{M. L\'{o}pez de Haro\cite{mariano} and J.A. del 
R\'{\i}o\cite{antonio} \\
%EndAName
Centro de Investigaci\'on en Energ\'{\i}a, UNAM,\\
Temixco, Morelos 62580, M\'{e}xico \and F. V\'{a}zquez\cite{federico} \\
%EndAName
Facultad de Ciencias,\\
Universidad Aut\'{o}noma del Estado de Morelos,\\
Cuernavaca, Mor. M\'{e}xico}
\date{\today}
\maketitle

\begin{abstract}
The light-scattering spectrum for a Maxwell fluid in a steady state due to
the presence of an external temperature gradient is computed. In such a
fluid heat conduction is assumed to be governed by the classical Fourier
law. The calculation is carried out through the use of fluctuating
hydrodynamics. The effect of the non-Newtonian character of the fluid in the
resulting spectrum is discussed.

\end{abstract}

\baselineskip 0.9 cm

\section{INTRODUCTION.}

When light passes through a fluid, correlations over space and time in the
density fluctuations are responsible for the scattering of light. In fact,
the dynamic structure factor $S(\mathbf{k,}\omega )$, which is the
space-time Fourier transform of the space-time correlations of density
fluctuations in the fluid\cite{BP}, reflects the strength of the exchanges
of energy and momentum between a light beam and the fluid as a function of
wavevector $\mathbf{k}$ and frequency $\omega $. For a one-component
Newtonian fluid in equilibrium, the theory for $S(\mathbf{k,}\omega )$ was
first given correctly in 1934 by Landau and Placzek\cite{LP} and later
rederived using fluctuating hydrodynamics\cite{LL} and the approach
advocated by Mountain \cite{mountain66},\cite{mountain66bis},\cite
{mountain68}. In fluctuating hydrodynamics, random fluxes are added to the
hydrodynamic equations (linearized about equilibrium) which then appear as
stochastic equations of the Langevin type. The fluctuation-dissipation
theorem provides a connection between the spectrum of spontaneous
fluctuations in thermal equilibrium and the transport properties of the
system. In contrast, the dynamic structure factor in Mountain's work is
derived directly from the linearized hydrodynamic equations without
resorting to any kind of fluctuating term added to the constitutive
equations. It is worthwhile to note here that both routes lead to the same
results and that these results have been confirmed many times by light
scattering measurements in various systems.

For nonequilibrium fluctuations in Newtonian fluids, the theory was
developed rather more recently\cite{NonEq1} - \cite{NonEq2} and the few
available experiments seem also to confirm its validity\cite{Beysens}-\cite
{Jan2}. The extension of fluctuating hydrodynamics to steady states is made
on the basis of the simple hypothesis that the fast degrees of freedom
behave as ``if they were in local equilibrium at the value of the
thermodynamic parameters determined from the solution to the steady state
problem'' \cite{tremb81}. With this statement it is assumed that the
statistics of the stochastic forces behaves as if they were within a local
equilibrium environment around each point \textbf{r} of the system \cite
{trembrecent}. It is usual under this hypothesis to rewrite the correlation
functions of the stochastic fluxes by substituting the equilibrium
thermodynamic properties and transport coefficients by their space dependent
values. Another way to see this assumption is based on the Markov property:
the random fluxes are not affected by the applied gradients since the
correlation length is of microscopic order \cite{NonEq2}.

Studies of fluctuations in non-Newtonian fluids are also relatively
abundant. Some of these systems have also been described under the scope of
linear Langevin approaches, as is the case of viscoelastic fluids near
equilibrium states \cite{Rytov}, suspensions in viscoelastic fluids under
density and velocity gradients \cite{bonet90} and colloidal suspensions
including memory effects \cite{NonEq2}. On the other hand, there are many
treatments \textit{\`{a} la Mountain }in this kind of fluids as well \cite
{Visc1} - \cite{Visc2}. Among these, although at the time meant in a
different context, we want to mention here explicitly the study of
(equilibrium) Rayleigh-Brillouin scattering in polyatomic fluids within an
extended irreversible thermodynamic treatment \cite{rodriguez87} carried out
by L\'{o}pez de Haro, Rodr\'{i}guez and Garc\'{i}a-Col\'{i}n a few years ago 
\cite{lopez87}. There, no explicit emphasis on the non-Newtonian character
of the fluid was made, although of course the evolution equations for the
trace and the symmetric traceless part of the stress tensor clearly imply
this character. More recently, V\'{a}zquez and L\'{o}pez de Haro\cite{Visc2}
have shown the equivalence of the computation of the equilibrium dynamic
structure factor of a Maxwell fluid both using fluctuating hydrodynamics and
Mountain's approach. It is a major aim of the present paper to profit from
the results of these previous works and to generalize them to the case where
the fluid is subjected to an external temperature gradient. The idea is to
assess to what extent non-Newtonian effects, which have been considered
negligible in actual experiments, do affect the light-scattering spectrum.

The paper is organized as follows. In Sect. II we compute the equilibrium
dynamic structure factor of a Maxwell fluid in which thermal conduction is
governed by Fourier's law. This is done using fluctuating hydrodynamics.
Sect. III\ deals with the same fluid which has now been brought to a steady
state by the action of an externally imposed temperature gradient. Here we
compute the light-scattering spectrum with the aid of the extension of
fluctuating hydrodynamics to nonequilibrium steady states. The paper is
closed in Sect. IV\ with a discussion and some concluding remarks.

\section{DYNAMIC\ STRUCTURE\ FACTOR\ FOR\ A\ MAXWELL\ FLUID\ IN EQUILIBRIUM.}

We here first analyze the case of a Maxwell fluid in equilibrium (\textit{%
i.e. }a fluid in which instead of the Navier-Newton law, both elastic and
viscous effects are incorporated into the constitutive equation for the
stress tensor, c.f. Eq. (\ref{27}) below) and we further assume that heat
conduction is governed by the classical Fourier law. For this case, the
corresponding \ full hydrodynamic equations (written in terms of the
density, velocity and entropy density of the fluid denoted by $\rho ,\mathbf{%
v}$ and $s$, respectively) are the continuity equation, the momentum balance
equation and the energy balance equation, namely

\begin{equation}
\frac{\partial \rho }{\partial t}+\mathbf{v}\cdot \nabla \rho =-\rho \nabla
\cdot \mathbf{v,}  \label{1}
\end{equation}

\begin{equation}
\rho \frac{\partial \mathbf{v}}{\partial t}+\rho \mathbf{v\cdot }\nabla 
\mathbf{v=-}\nabla p-\nabla \cdot \stackrel{\leftrightarrow }{\tau },
\label{25}
\end{equation}
and

\begin{equation}
\rho T\frac{\partial s}{\partial t}+\rho T\mathbf{v\cdot }\nabla s=-\nabla
\cdot \mathbf{q}-\stackrel{\leftrightarrow }{\tau }:\nabla \mathbf{v,}
\label{26}
\end{equation}
which must be complemented with the constitutive equations for the heat flux
vector $\mathbf{q}$ and the stress tensor $\stackrel{\leftrightarrow }{\tau }
$ . These equations in our case read

\begin{equation}
\mathbf{q}=-\lambda \nabla T  \label{27a}
\end{equation}
and 
\begin{equation}
-\tau _{r}\frac{\partial \stackrel{\leftrightarrow }{\tau }}{\partial t}%
-\tau _{r}\mathbf{v\cdot }\nabla \stackrel{\leftrightarrow }{\tau }=%
\stackrel{\leftrightarrow }{\tau }+2\eta \stackrel{\circ }{\nabla \mathbf{v}}%
+\eta _{v}\left( \nabla \cdot \mathbf{v}\right) \stackrel{\leftrightarrow }{1%
}.  \label{27}
\end{equation}
Here, $p$ is the hydrostatic pressure, $T$ is the temperature, $\eta $ is
the shear viscosity, $\eta _{v}$ is the bulk viscosity, $C_{\rho }$ is the
constant density heat capacity, $\lambda $ is the thermal conductivity, $%
\stackrel{\leftrightarrow }{1}$ is the unit tensor, $\stackrel{\circ }{%
\nabla \mathbf{v}}=\frac{1}{2}\left[ \nabla \mathbf{v+(}\nabla \mathbf{v)}%
^{T}\right] -\frac{1}{3}\nabla \cdot \mathbf{v}\stackrel{\leftrightarrow }{1}
$ stands for the symmetric and traceless part of the velocity gradient and $%
\tau _{r}$ is the Maxwell relaxation time. In Eq. (\ref{25}) one can
eliminate the hydrostatic pressure in favor of $\rho $ and $s$ using the
fact that $\nabla p=$ $\mathbf{-}c^{2}\nabla \rho +\frac{T\alpha }{C_{\rho
}\kappa _{T}}\nabla s$ (where $c$ is the adiabatic speed of sound in the
fluid, $\alpha $ is the thermal expansion coefficient and $\kappa _{T}$ is
the isothermal compressibility) but the form we have chosen will prove
convenient later on. Note \ also that on the left hand side of Eq. (\ref{27}%
) we have considered the usual material time derivative rather than the
corrotational time derivative. Although the latter choice would be more
satisfactory from the rheological point of view, Eq. (\ref{27}) represents
the simplest form of the Maxwell model. In any case, since later on we will
retain only the linear terms, the use of the material time derivative will
not imply loss of generality.

We consider for simplicity the case where $\alpha =0$ (which implies that $%
C_{\rho }=C_{p}$, where $C_{p}$ is the constant pressure heat capacity) and
take $\eta ,$ $\eta _{v\text{ }},$ $c,$ $\lambda ,$ $C_{\rho }$ and $\tau
_{r}$ to be constant. Let a subscript zero refer to the equilibrium value
and a subscript one denote the (small) deviation from equilibrium, so that $%
\rho =\rho _{0}+\rho _{1}$, $\mathbf{v}=\mathbf{v}_{1}$, $s=s_{0}+s_{1}$, $%
T=T_{0}+T_{1}$and $\stackrel{\leftrightarrow }{\tau }=\stackrel{%
\leftrightarrow }{\tau }_{1}$. Keeping only linear contributions in the
deviations, one gets from Eqs. (\ref{1}), \ and (\ref{25} ) to (\ref{27})

\begin{equation}
\frac{\partial \rho _{1}}{\partial t}=-\rho _{0}\nabla \cdot \mathbf{v}_{1},
\label{28}
\end{equation}

\begin{equation}
\rho _{0}\frac{\partial \mathbf{v}_{1}}{\partial t}\mathbf{=-}c^{2}\nabla
\rho _{1}-\nabla \cdot \stackrel{\leftrightarrow }{\tau }_{1},  \label{29}
\end{equation}

\begin{equation}
\rho _{0}T_{0}\frac{\partial s_{1}}{\partial t}=\lambda \nabla ^{2}T_{1},
\label{30}
\end{equation}
and

\begin{equation}
-\tau _{r}\frac{\partial \stackrel{\leftrightarrow }{\tau }_{1}}{\partial t}%
= \text{ }\stackrel{\leftrightarrow }{\tau }_{1}+2\eta \stackrel{\circ }{
\nabla \mathbf{v}_{1}}+\eta _{v}\left( \nabla \cdot \mathbf{v}_{1}\right) 
\stackrel{\leftrightarrow }{1}.  \label{31}
\end{equation}

Within fluctuating hydrodynamics\cite{LL}, stochastic contributions $%
\stackrel{\leftrightarrow }{\sigma }$ and $\mathbf{g}$ are added to the
momentum and heat fluxes, respectively. At this stage it is convenient to
work in Fourier space. Denoting by an upper tilde the space and time Fourier
transform of a given quantity, the linearized continuity, momentum and
energy equations in $\left( \mathbf{k},\omega \right) $ - space (with $%
\mathbf{k}$ the wavevector and $\omega $ the frequency) are written as

\begin{equation}
-i\omega \widetilde{\rho }_{1}=-i\rho _{0}\mathbf{k\cdot }\widetilde{\mathbf{%
v}} _{1},  \label{32}
\end{equation}

\begin{equation}
-i\omega \rho _{0}\widetilde{\mathbf{v}}_{1}=-ic^{2}\mathbf{k}\widetilde{%
\rho }_{1}-\frac{\left( 1+i\omega \tau _{r}\right) }{1+\omega ^{2}\tau
_{r}^{2}}\left( \frac{4}{3}\eta +\eta _{v}\right) k^{2}\widetilde{\mathbf{v}}%
_{1}-i\mathbf{k\cdot }\stackrel{\leftrightarrow }{\widetilde{\sigma }},
\label{33}
\end{equation}
and

\begin{equation}
-i\omega \rho _{0}T_{0}\widetilde{s}_{1}=-\lambda k^{2}\widetilde{T}_{1}-i%
\mathbf{k\cdot }\widetilde{\mathbf{g}}.  \label{34}
\end{equation}
Note that in writing Eq. (\ref{33}) we have made use of the $\left( \mathbf{k%
},\omega \right) $ - space version of Eq. (\ref{31}), namely

\begin{equation}
\stackrel{\leftrightarrow }{\widetilde{\tau }}_{1}=-\frac{\left( 1+i\omega
\tau _{r}\right) }{1+\omega ^{2}\tau _{r}^{2}}\left[ 2i\eta \stackrel{\circ 
}{(\mathbf{k}\widetilde{\mathbf{v}}_{1})}+i\eta _{v}\mathbf{k\cdot }%
\widetilde{\mathbf{v} }_{1}\stackrel{\leftrightarrow }{1}\right] .
\label{35}
\end{equation}

\bigskip

\bigskip In order to close the fluctuating hydrodynamics approach, we now
require the corresponding fluctuation-dissipation relations. Since in the $(%
\mathbf{r},\omega )$- space it follows from Eq. (\ref{31}) that

\begin{eqnarray}
\stackrel{\leftrightarrow }{\tau }_{1}\left( \mathbf{r},\omega \right) &=&-%
\frac{\left( 1+i\omega \tau _{r}\right) }{1+\omega ^{2}\tau _{r}^{2}}\left[
2\eta \stackrel{\circ }{\nabla \mathbf{v}_{1}}\left( \mathbf{r},\omega
\right) +\eta _{v}\left( \nabla \cdot \mathbf{v}_{1}\left( \mathbf{r},\omega
\right) \right) \stackrel{\leftrightarrow }{1}\right]  \nonumber \\
&\equiv &2\eta ^{D}(\omega )\stackrel{\circ }{\nabla \mathbf{v}_{1}}\left( 
\mathbf{r},\omega \right) +\eta _{v}^{D}(\omega )\left( \nabla \cdot \mathbf{%
v}_{1}\left( \mathbf{r},\omega \right) \right) \stackrel{\leftrightarrow }{1}%
,  \label{stressw}
\end{eqnarray}
\textit{i.e. }in the frequency domain the Maxwell model implies a\textit{\
linear} relationship between the stress tensor and the thermodynamic forces
but with dispersive transport coefficients, $\eta ^{D}(\omega )$ and $\eta
_{v}^{D}(\omega )$ being the dispersive (complex and frequency-dependent)
transport coefficients, then following Landau and Lifshitz \cite{LL} the
fluctuation-dissipation relations read

\begin{equation}
\left\langle \widetilde{g}_{i}\left( \mathbf{k,}\omega \right) \widetilde{g}%
_{j}^{\ast }(\mathbf{k}^{\prime },\omega ^{\prime })\right\rangle
=2k_{B}T_{0}^{2}\left( 2\pi \right) ^{4}\lambda \delta _{ij}\delta
^{3}\left( \mathbf{k-k}^{\prime }\right) \delta \left( \omega -\omega
^{\prime }\right) ,  \label{11}
\end{equation}

\begin{equation}
\left\langle \widetilde{\sigma }_{ij}\left( \mathbf{k,}\omega \right) 
\widetilde{\sigma }_{lm}^{\ast }(\mathbf{k}^{\prime },\omega ^{\prime
})\right\rangle =\frac{2k_{B}T_{0}}{1+\omega ^{2}\tau _{r}^{2}}\left( 2\pi
\right) ^{4}\left[ 2\eta \Delta _{ijlm}\mathbf{+}\left( \eta _{v}-\frac{2}{3}%
\eta \right) \delta _{ij}\delta _{lm}\right] \delta ^{3}\left( \mathbf{k-k}%
^{\prime }\right) \delta \left( \omega -\omega ^{\prime }\right) .
\label{40bb}
\end{equation}
and

\begin{equation}
\left\langle \widetilde{\sigma }_{ij}\left( \mathbf{k,}\omega \right) 
\widetilde{g}_{l}^{\ast }(\mathbf{k}^{\prime },\omega ^{\prime
})\right\rangle =\left\langle \widetilde{\sigma }_{ij}\left( \mathbf{k,}%
\omega \right) \right\rangle =\left\langle \widetilde{g}_{l}(\mathbf{k}%
,\omega )\right\rangle =0  \label{11a}
\end{equation}

\noindent the angular brackets denoting an (equilibrium) ensemble average
over the initial states of the system, the asterisk meaning complex
conjugation and $\Delta _{ijlm}=\frac{1}{2}\left( \delta _{il}\delta
_{jm}+\delta _{im}\delta _{jl}\right) $, $\delta _{rs}$ being the Kronecker
delta.

Note that the energy equation within the present approximation is decoupled
from both the continuity equation and the momentum balance equation. Since
the dynamic structure factor $S(\mathbf{k,}\omega )$ is related to the
density-density correlation function through

\begin{equation}
\left\langle \widetilde{\rho }_{1}\left( \mathbf{k,}\omega \right) 
\widetilde{\rho }_{1}^{\ast }\left( \mathbf{k}^{\prime }\mathbf{,}\omega
^{\prime }\right) \right\rangle \equiv \left( 2\pi \right) ^{4}\delta
^{3}\left( \mathbf{k-k}^{\prime }\right) \delta \left( \omega -\omega
^{\prime }\right) S_{B}^{eq}\left( \mathbf{k,}\omega \right) ,  \label{12}
\end{equation}

\noindent the subscript $B$ having been introduced to stress the fact that,
within the present approximation, it will give the Brillouin spectrum, and
since from Eqs. (\ref{32}), (\ref{33}) and (\ref{40bb}) it follows that

\[
\left\langle \widetilde{\rho }_{1}\left( \mathbf{k,}\omega \right) 
\widetilde{\rho }_{1}^{\ast }\left( \mathbf{k}^{\prime }\mathbf{,}\omega
^{\prime }\right) \right\rangle =\left( 2\pi \right) ^{4}\delta ^{3}\left( 
\mathbf{k-k}^{\prime }\right) \delta \left( \omega -\omega ^{\prime }\right) 
\]
\begin{equation}
\times \frac{2k_{B}T_{0}\left( 1+\omega ^{2}\tau _{r}^{2}\right) \left( 
\frac{4}{3}\eta +\eta _{v}\right) k^{4}}{\left[ \left( \omega
^{2}-c^{2}k^{2}\right) \left( 1+\omega ^{2}\tau _{r}^{2}\right) -\omega ^{2}%
\frac{\tau _{r}}{\rho _{0}}\left( \frac{4}{3}\eta +\eta _{v}\right) k^{2}%
\right] ^{2}+\frac{\omega ^{2}}{\rho _{0}^{2}}\left( \frac{4}{3}\eta +\eta
_{v}\right) ^{2}k^{4}},  \label{13}
\end{equation}

\noindent one readily identifies

\begin{equation}
S_{B}^{eq}\left( \mathbf{k,}\omega \right) =\frac{2k_{B}T_{0}\left( 1+\omega
^{2}\tau _{r}^{2}\right) \left( \frac{4}{3}\eta +\eta _{v}\right) k^{4}}{%
\left[ \left( \omega ^{2}-c^{2}k^{2}\right) \left( 1+\omega ^{2}\tau
_{r}^{2}\right) -\omega ^{2}\frac{\tau _{r}}{\rho _{0}}\left( \frac{4}{3}%
\eta +\eta _{v}\right) k^{2}\right] ^{2}+\frac{\omega ^{2}}{\rho _{0}^{2}}%
\left( \frac{4}{3}\eta +\eta _{v}\right) ^{2}k^{4}}.  \label{14}
\end{equation}

The spectrum given in Eq. (\ref{14}) consists of two (symmetric) Brillouin
peaks slightly displaced from $\omega _{\pm }=\pm ck$ and a peak centered at 
$\omega _{c}=0$, known as the Mountain peak, whose origin can be ascribed to
the viscoelastic character of the fluid\cite{MP}.

Although in the approximation we have taken of $\alpha =0$, there is no
Rayleigh line, if this restriction were avoided, the Rayleigh line, also
centered around $\omega _{c}=0$, would follow from the entropy-entropy
correlation function. In any case, since to a first approximation the
Rayleigh line is determined by the coupling of the temperature fluctuations
to the transverse velocity fluctuations\cite{Jan1}, the missing terms
(proportional to $\alpha $) in Eqs. (\ref{33}) and (\ref{34}) will play no
role in the final result. So we compute the entropy-entropy correlation
function by manipulating Eqs. (\ref{34}) and (\ref{11}). The result is

\begin{equation}
\left\langle \widetilde{s}_{1}\left( \mathbf{k},\omega \right) \widetilde{s}%
_{1}^{\ast }\left( \mathbf{k},\omega \right) \right\rangle =S_{R}^{eq}\left( 
\mathbf{k,}\omega \right) =\frac{2\left( 2\pi \right)
^{4}k_{B}T_{0}^{2}C_{p}D_{T}k^{2}}{\rho _{0}\left( \omega
^{2}+D_{T}^{2}k^{4}\right) },  \label{RAYE}
\end{equation}
where $D_{T}=\frac{\lambda }{\rho _{0C_{p}}}$ is the thermal diffusivity and
the subscript $R$ serves to make reference to the connection with the
Rayleigh line.

\section{FLUCTUATING\ HYDRODYNAMICS\ \ AND\ THE\ LIGHT-SCATTERING SPECTRUM\
OF\ A\ MAXWELL\ FLUID\ UNDER\ AN\ EXTERNAL\ TEMPERATURE\ GRADIENT.}

We now examine the same Maxwell fluid in which heat conduction is governed
by Fourier%
%TCIMACRO{\UNICODE{0xb4}}%
%BeginExpansion
\'{}%
%EndExpansion
s law but in a layer confined between two parallel plates at $z=+\frac{L}{2}$
and $z=-\frac{L}{2}$ and of infinite extension in the $xy-$plane. In the
fluid layer a steady temperature gradient is maintained by fixing the
temperatures of the upper and lower planes at different values, namely $%
T_{1} $ at $z=-\frac{L}{2}$ and $T_{2}$ at $z=+\frac{L}{2}$. We assume that
neither the gravitational force nor any other external forces act on the
system. A time-independent solution to Eqs. (\ref{1}) to (\ref{27}) leading
to a stationary state (denoted by a subscript $s$) for the system satisfies
the relations

\begin{equation}
\mathbf{v}_{s}=\mathbf{0}  \label{FE1}
\end{equation}
\begin{equation}
\nabla p_{s}=0  \label{FE2}
\end{equation}
\begin{equation}
\nabla \cdot \mathbf{q}_{s}=0.  \label{FE3}
\end{equation}
Due to the symmetry of the problem, Eq. (\ref{FE3}) together with Eq.(\ref
{27a}) leads to

\begin{equation}
\frac{d}{dz}\left( \lambda _{s}\frac{dT_{s}}{dz}\right) =\left( \frac{%
\partial \lambda }{\partial T}\right) _{s}\left( \frac{dT_{s}}{dz}\right)
^{2}+\lambda _{s}\frac{d^{2}T}{dz^{2}}=0.  \label{temp1}
\end{equation}

For simplicity, let us look at the case where $\alpha _{s}=0$ and take $\eta
_{s},$ $\eta _{vs\text{ }},$ $c_{s},$ $\lambda _{s},$ $C_{ps}$ and $\tau
_{rs}$ to be constant. Since $\lambda _{s}$ has been assumed to be a
constant, it follows from Eq. (\ref{temp1}) \ and the boundary conditions
that 
\[
T_{s}=T_{0}-Az 
\]
where 
\[
T_{0}=\frac{T_{1}+T_{2}}{2}=T_{s}(0) 
\]
and 
\[
A=\frac{T_{1}-T_{2}}{L}. 
\]

Further, the condition implied by Eq. (\ref{FE2}) leads to $\nabla \rho
_{s}=0$ or $\rho _{s}=\rho _{0}$, while Eq. (\ref{FE1}) implies that $%
\stackrel{\leftrightarrow }{\tau }_{s}=\stackrel{\leftrightarrow }{0}$. Let
a subscript $1s$ now denote the (small) deviation from the steady state, so
that $\rho =\rho _{s}+\rho _{1s}$, $\mathbf{v}=\mathbf{v}_{s}+\mathbf{v}%
_{1s}=\mathbf{v}_{1s}$, $s=s_{s}+s_{1s}$, $T=T_{s}+T_{1s}$, $\mathbf{q}=%
\mathbf{q}_{s}+\mathbf{q}_{1s}=-\lambda _{s}\nabla T_{s}+\mathbf{q}_{1s}$
and $\stackrel{\leftrightarrow }{\tau }=\stackrel{\leftrightarrow }{\tau }%
_{s}+\stackrel{\leftrightarrow }{\tau }_{1s}=\stackrel{\leftrightarrow }{%
\tau }_{1s}$. Keeping only linear contributions in the deviations, one gets
from Eqs. (\ref{1}) to (\ref{27})

\begin{equation}
\frac{\partial \rho _{1s}}{\partial t}=-\rho _{s}\nabla \cdot \mathbf{v}%
_{1s},  \label{28bis}
\end{equation}

\begin{equation}
\rho _{s}\frac{\partial \mathbf{v}_{1s}}{\partial t}\mathbf{=-}%
c_{s}^{2}\nabla \rho _{1s}-\nabla \cdot \stackrel{\leftrightarrow }{\tau }%
_{1s},  \label{29bis}
\end{equation}

\begin{equation}
\rho _{0}T_{0}\frac{\partial s_{1s}}{\partial t}+\rho _{0}T_{0}\mathbf{v}%
_{1s}\cdot \nabla T_{s}=-\nabla \cdot \mathbf{q}_{1s},  \label{30bis}
\end{equation}

\begin{equation}
\mathbf{q}_{1s}=-\lambda _{s}\nabla T_{1s},  \label{30abis}
\end{equation}
and

\begin{equation}
-\tau _{rs}\frac{\partial \stackrel{\leftrightarrow }{\tau }_{1s}}{\partial t%
}=\text{ }\stackrel{\leftrightarrow }{\tau }_{1s}+2\eta _{s}\stackrel{\circ 
}{\nabla \mathbf{v}_{1s}}+\eta _{vs}\left( \nabla \cdot \mathbf{v}%
_{1s}\right) \stackrel{\leftrightarrow }{1}.  \label{31bis}
\end{equation}

The balance equations in the $(\mathbf{r},\omega )-$ domain read

\begin{equation}
-i\omega \rho _{1s}=-\rho _{s}\nabla \cdot \mathbf{v}_{1s},  \label{32bis}
\end{equation}

\begin{equation}
-i\omega \rho _{s}\mathbf{v}_{1s}=\mathbf{-}c_{s}^{2}\nabla \rho
_{1s}-\nabla \cdot \stackrel{\leftrightarrow }{\tau }_{1s},  \label{33bis}
\end{equation}
and

\begin{equation}
-i\omega \rho _{0}T_{0}s_{1s}+A\rho _{0}C_{ps}\mathbf{v}_{1s}\cdot \widehat{%
\mathbf{z}}=-\nabla \cdot \mathbf{q}_{1s}.  \label{34bis}
\end{equation}
where $\widehat{\mathbf{z}}$ is a unit vector in the $z$ direction. We now
add fluctuating contributions to the heat flux and stress tensor denoted by $%
\mathbf{g}_{s}$ and $\stackrel{\leftrightarrow }{\sigma }_{s}$,
respectively, take the spatial Fourier transform and make use of the
stationary state analog of Eq. (\ref{stressw}) to arrive at

\begin{equation}
-i\omega \widetilde{\rho }_{1s}=-i\rho _{s}\mathbf{k}\cdot \widetilde{%
\mathbf{v}}_{1s},  \label{35bis}
\end{equation}

\begin{equation}
-i\omega \rho _{s}\widetilde{\mathbf{v}}_{1s}=-ic_{s}^{2}\mathbf{k}%
\widetilde{\rho }_{1s}-\frac{\left( 4/3\eta _{s}+\eta _{vs}\right) \left(
1+i\omega \tau _{rs}\right) k^{2}}{\left( 1+\omega ^{2}t_{rs}^{2}\right) }%
\widetilde{\mathbf{v}}_{1s}-i\mathbf{k}\cdot \widetilde{\stackrel{%
\leftrightarrow }{\sigma }}_{s},  \label{36bis}
\end{equation}
and

\begin{equation}
-i\omega \rho _{0}T_{0}\widetilde{s}_{1s}=-\rho _{0}T_{0}D_{Ts}k^{2}%
\widetilde{s}_{1s}-i\mathbf{k}\cdot \widetilde{\mathbf{g}}_{s}-A\rho
_{0}C_{ps}\mathbf{v}_{1s}\cdot \widehat{\mathbf{z}}.  \label{37bis}
\end{equation}
\qquad\ 

In order to proceed any further, we require the fluctuation-dissipation
relations obeyed by $\widetilde{\mathbf{g}}_{s}$ and $\widetilde{\stackrel{%
\leftrightarrow }{\sigma }}_{s}$. Under the hypothesis mentioned in the
Introduction, we follow previous authors in assuming that the
fluctuation-dissipation relations have the same form as that for thermal
equilibrium, but with the equilibrium temperature replaced by the stationary
temperature and the average taken over an stationary ensemble. Thus,
restricting to small gradients (linear order in $A$) in our case one gets 
\[
\left\langle \widetilde{g}_{is}\left( \mathbf{k,}\omega \right) \widetilde{g}%
_{js}^{\ast }(\mathbf{k}^{\prime },\omega ^{\prime })\right\rangle
=2k_{B}T_{0}^{2}\left( 2\pi \right) ^{4}\lambda _{s}\delta _{ij}\delta
\left( \omega -\omega ^{\prime }\right) 
\]
\begin{equation}
\times \left[ \delta ^{3}\left( \mathbf{k-k}^{\prime }\right) +\frac{iAL}{%
T_{0}}\left( \delta ^{3}\left( \mathbf{k-k}^{\prime }+\widehat{\mathbf{z}}%
\right) -\delta ^{3}\left( \mathbf{k-k}^{\prime }-\frac{\widehat{\mathbf{z}}%
}{L}\right) \right) \right] ,  \label{38bis}
\end{equation}
\[
\left\langle \widetilde{\sigma }_{sij}\left( \mathbf{k,}\omega \right) 
\widetilde{\sigma }_{slm}^{\ast }(\mathbf{k}^{\prime },\omega ^{\prime
})\right\rangle =\frac{2k_{B}T_{0}}{1+\omega ^{2}\tau _{r}^{2}}\left( 2\pi
\right) ^{4}\left[ 2\eta _{s}\Delta _{ijlm}\mathbf{+}\left( \eta _{vs}-\frac{%
2}{3}\eta _{s}\right) \delta _{ij}\delta _{lm}\right] \delta \left( \omega
-\omega ^{\prime }\right) 
\]

\begin{equation}
\times \left[ \delta ^{3}\left( \mathbf{k-k}^{\prime }\right) +\frac{iAL}{%
2T_{0}}\left( \delta ^{3}\left( \mathbf{k-k}^{\prime }+\frac{\widehat{%
\mathbf{z}}}{L}\right) -\delta ^{3}\left( \mathbf{k-k}^{\prime }-\frac{%
\widehat{\mathbf{z}}}{L}\right) \right) \right] .  \label{39bis}
\end{equation}
while

\begin{equation}
\left\langle \widetilde{\sigma }_{sij}\left( \mathbf{k,}\omega \right) 
\widetilde{g}_{sl}^{\ast }(\mathbf{k}^{\prime },\omega ^{\prime
})\right\rangle =\left\langle \widetilde{\sigma }_{sij}\left( \mathbf{k,}%
\omega \right) \right\rangle =\left\langle \widetilde{g}_{sl}(\mathbf{k}%
,\omega )\right\rangle =0.  \label{40bis}
\end{equation}

Tremblay, Arai and Siggia\cite{tremb81} have shown that, in a light
scattering experiment in which a fluid in a stationary state due to an
external temperature gradient is involved, 
\begin{eqnarray*}
S_{B}\left( \mathbf{k},\omega \right) &\equiv &\left\langle \widetilde{\rho }%
_{1}\left( \mathbf{k},\omega \right) \widetilde{\rho }_{1}^{\ast }\left( 
\mathbf{k},\omega \right) \right\rangle +\left\langle \widetilde{\rho }%
_{1}\left( \mathbf{k+}\widehat{\mathbf{z}}/2L,\omega \right) \widetilde{\rho 
}_{1}^{\ast }\left( \mathbf{k-}\widehat{\mathbf{z}}/2L,\omega \right)
\right\rangle \\
&&+\left\langle \widetilde{\rho }_{1}\left( \mathbf{k-}\widehat{\mathbf{z}}%
/2L,\omega \right) \widetilde{\rho }_{1}^{\ast }(\mathbf{k+}\widehat{\mathbf{%
z}}/2L,\omega )\right\rangle
\end{eqnarray*}
is the correlation function of interest for the Brillouin spectrum. By a
simple manipulation of Eqs. (\ref{35bis}), (\ref{36bis}) and (\ref{39bis})
it is easy to arrive at

\[
S_{B}\left( \mathbf{k},\omega \right) =S_{B}^{eq}\left( \mathbf{k},\omega
\right) \left[ 1-\frac{2\omega ^{3}D_{ls}\left( \omega \right) A\left( 
\mathbf{k}\cdot \widehat{\mathbf{z}}\right) }{T_{0}\left[ \left( \omega
^{2}-C_{s}^{2}\left( \omega \right) k^{2}\right) ^{2}+\left( \omega
D_{ls}\left( \omega \right) k^{2}\right) ^{2}\right] }\right] 
\]

\begin{equation}
\simeq \frac{k_{B}T_{0}\rho _{0}}{C_{s}^{2}\left( \omega \right) }\frac{%
D_{ls}\left( \omega \right) k^{2}}{2}\left[ \frac{1-\varepsilon \left( 
\mathbf{k},\omega \right) }{\left( \omega -C_{s}\left( \omega \right)
k\right) ^{2}+\frac{1}{4}\left( D_{ls}\left( \omega \right) k^{2}\right) ^{2}%
}+\frac{1+\varepsilon \left( \mathbf{k},\omega \right) }{\left( \omega
+C_{s}\left( \omega \right) k\right) ^{2}+\frac{1}{4}\left( D_{ls}\left(
\omega \right) k^{2}\right) ^{2}}\right] ,  \label{41bis}
\end{equation}
where $S_{B}^{eq}\left( \mathbf{k},\omega \right) $ \ is given by Eq. (\ref
{14}) and we have identified $\rho _{s}=\rho _{0},\eta _{s}=\eta ,$ $\eta
_{vs\text{ }}=\eta _{v\text{ }},$ $c_{_{s}}=c,$ $\ $and $\tau _{rs}=\tau
_{r} $ and introduced the following quantities

\begin{equation}
D_{ls}\left( \omega \right) =\frac{\frac{4}{3}\eta +\eta _{v}}{\rho
_{0}\left( 1+\omega ^{2}\tau _{r}^{2}\right) },  \label{42bis}
\end{equation}
\begin{equation}
C_{s}\left( \omega \right) =\left[ c^{2}+\omega ^{2}\tau _{r}D_{ls}\left(
\omega \right) \right] ^{1/2},  \label{43bis}
\end{equation}
and 
\begin{equation}
\varepsilon \left( \mathbf{k},\omega \right) =\frac{C_{s}\left( \omega
\right) AL\left( \mathbf{k}\cdot \widehat{\mathbf{z}}\right) }{%
T_{0}D_{ls}\left( \omega \right) k^{2}}\left[ \frac{2\left( \omega
D_{ls}\left( \omega \right) k^{2}\right) ^{2}}{\left( \omega
^{2}-C_{s}^{2}\left( \omega \right) k^{2}\right) ^{2}+\left( \omega
D_{ls}\left( \omega \right) k^{2}\right) ^{2}}\right] .  \label{44bis}
\end{equation}

It should be pointed out that in order to perform the expansion leading to
the approximation appearing in the second line of Eq. (\ref{41bis}), apart
from the weak gradient assumption (\textit{i. e.} $\frac{\left| AL\right| }{%
T_{0}}\ll 1$) it has also been assumed that \ $\left| \frac{C_{s}(\omega
)\left( \mathbf{k}\cdot \widehat{\mathbf{z}}\right) }{D_{ls}\left( \omega
\right) k^{2}}\right| \ll 1$.

Now we turn to the Rayleigh line. Again the restriction $\alpha _{s}=0$
implies that the spectrum in Eq. (\ref{41bis}) contains no Rayleigh line. If
such restriction is removed, in order to compute the Rayleigh line one again
requires the entropy-entropy correlation function. By the same line of
reasoning used in connection with Eq. (\ref{RAYE}), the missing terms
(proportional to $\alpha _{s}$) in Eqs. (\ref{36bis}) and (\ref{37bis}) will
play no role in the final result. So we proceed with the calculation. From
Eq. (\ref{37bis}) \ and making the identification $D_{Ts}=D_{T}$ it follows
that

\begin{equation}
\widetilde{s}_{1s}\left( \mathbf{k},\omega \right) =\frac{-i\mathbf{k}\cdot 
\widetilde{\mathbf{g}}_{s}/(\rho _{0}T_{0})-AC_{p}\widetilde{\mathbf{v}}%
_{1s}\cdot \widehat{\mathbf{z}}/T_{0}}{-i\omega +D_{T}k^{2}}.  \label{ttemp}
\end{equation}
We only need the transverse velocity fluctuations \ $\widetilde{v}_{ts}$
which are obtained from $\widetilde{\mathbf{v}}_{1s}$ as

\begin{equation}
\widetilde{v}_{ts}=\widetilde{\mathbf{v}}_{1s}\cdot \left( \stackrel{%
\leftrightarrow }{1}-\text{ }\widehat{\mathbf{k}}\widehat{\mathbf{k}}\right)
\cdot \widehat{\mathbf{z}}.  \label{velt}
\end{equation}

It is clear from this last equation that if the scattering vector $\mathbf{k}
$ is taken parallel to the temperature gradient, then $\widetilde{v}_{ts}=0$%
, so there will be no effect on the entropy-entropy correlation function.
Therefore we choose $\mathbf{k}\perp \widehat{\mathbf{z}}$. From Eqs. (\ref
{36bis}) and (\ref{stressw}) it follows that

\begin{equation}
\widetilde{v}_{ts}=-\frac{i\mathbf{k}\cdot \widetilde{\stackrel{%
\leftrightarrow }{\sigma }}_{s}\cdot \left( \stackrel{\leftrightarrow }{1}-%
\text{ }\widehat{\mathbf{k}}\widehat{\mathbf{k}}\right) \cdot \widehat{%
\mathbf{z}}}{-i\omega \rho _{0}+\frac{\eta \left( 1+i\omega \tau _{r}\right)
k^{2}}{\left( 1+\omega ^{2}t_{r}^{2}\right) }}.  \label{velt2}
\end{equation}

Using this expression in Eq. (\ref{ttemp}) yields

\begin{equation}
\widetilde{s}_{1s}\left( \mathbf{k},\omega \right) =\frac{-i\mathbf{k}\cdot 
\widetilde{\mathbf{g}}_{s}}{\rho _{0}T_{0}(-i\omega +D_{T}k^{2})}+\frac{%
iAC_{p}\mathbf{k}\cdot \widetilde{\stackrel{\leftrightarrow }{\sigma }}%
_{s}\cdot \left( \stackrel{\leftrightarrow }{1}-\text{ }\widehat{\mathbf{k}}%
\widehat{\mathbf{k}}\right) \cdot \widehat{\mathbf{z}}}{\rho
_{0}T_{0}(-i\omega +D_{T}k^{2})\left( -i\omega +\frac{\eta \left( 1+i\omega
\tau _{r}\right) k^{2}}{\rho _{0}\left( 1+\omega ^{2}t_{r}^{2}\right) }%
\right) }.  \label{entf}
\end{equation}

Since any asymmetries occur in the $\widehat{\mathbf{z}}$ direction, in the
scattering plane (which is orthogonal to the temperature gradient) spatial
and temporal invariance holds. This in turn implies that from the
fluctuation-dissipation relations one finds (for $\mathbf{k}$, $\mathbf{k}%
^{\prime }\perp \widehat{\mathbf{z}}$)

\begin{equation}
\left\langle \mathbf{k}\cdot \widetilde{\mathbf{g}}_{s}\left( \mathbf{k,}%
\omega \right) \mathbf{k}^{\prime }\cdot \widetilde{\mathbf{g}}_{s}^{\ast }(%
\mathbf{k}^{\prime },\omega ^{\prime })\right\rangle =2k_{B}T_{0}^{2}\left(
2\pi \right) ^{4}\rho _{0}C_{p}D_{T}k^{2}\delta ^{3}\left( 
\mathbf{k-k}^{\prime }\right) \delta \left( \omega -\omega ^{\prime }\right)
,  \label{TFDR1}
\end{equation}
and

\begin{equation}
\left\langle \mathbf{k}\cdot \widetilde{\stackrel{\leftrightarrow }{\sigma }}%
_{s}\cdot \left( \stackrel{\leftrightarrow }{1}-\text{ }\widehat{\mathbf{k}}%
\widehat{\mathbf{k}}\right) \cdot \widehat{\mathbf{z}}\cdot \mathbf{k}%
^{\prime }\cdot \widetilde{\stackrel{\leftrightarrow }{\sigma }}_{s}^{\ast
}\cdot \left( \stackrel{\leftrightarrow }{1}-\text{ }\widehat{\mathbf{k}}%
^{\prime }\widehat{\mathbf{k}}^{\prime }\right) \cdot \widehat{\mathbf{z}}%
\sigma \right\rangle =\frac{2k_{B}T_{0}}{1+\omega ^{2}\tau _{r}^{2}}\left(
2\pi \right) ^{4}\eta k^{2}\delta ^{3}\left( \mathbf{k-k}^{\prime }\right)
\delta \left( \omega -\omega ^{\prime }\right) .  \label{TFDR2}
\end{equation}

Use of Eqs. (\ref{entf}) to (\ref{TFDR2}) leads finally to 
\[
\left\langle \widetilde{s}_{1s}\left( \mathbf{k},\omega \right) \widetilde{s}%
_{1s}^{\ast }\left( \mathbf{k},\omega \right) \right\rangle \equiv
S_{R}\left( \mathbf{k},\omega \right) 
\]

\begin{equation}
=S_{R}^{eq}\left( \mathbf{k},\omega \right) \left[ 1+\frac{C_{p}\frac{\eta }{%
\rho _{0}\left( 1+\omega ^{2}\tau _{r}^{2}\right) }A^{2}}{T_{0}D_{T}\left[
\omega ^{2}\left( 1+\frac{\tau _{r}\eta k^{2}}{\rho _{0}\left( 1+\omega
^{2}\tau _{r}^{2}\right) }\right) ^{2}+\frac{\eta ^{2}k^{4}}{\rho
_{0}^{2}\left( 1+\omega ^{2}\tau _{r}^{2}\right) ^{2}}\right] }\right]
,\qquad (\mathbf{k}\perp \widehat{\mathbf{z}}),  \label{EEF}
\end{equation}
with $S_{R}^{eq}\left( \mathbf{k},\omega \right) $ given by Eq. (\ref{RAYE}%
). Eqs. (\ref{41bis}) and (\ref{EEF}) are the main results of this paper. A
brief assessment of their potential use will be carried out in the following
Section.

\section{DISCUSSION.}

As a first observation, we note that if the limit $\tau _{r}\rightarrow 0$
is taken, then Eqs. (\ref{14}), (\ref{RAYE}), (\ref{41bis}) and (\ref{EEF})
reduce to the well known results for the Newtonian fluid (\textit{c.f.} for
instance \ Refs. \cite{mountain66bis}, \cite{tremb81} and \cite{Jan1}), as
they should. Also, the equilibrium light-scattering spectrum in this paper,
Eq. (\ref{14}), arises as a particular case of the one computed in Ref. \cite
{lopez87} provided in the latter the relaxation times of the trace and
traceless symmetric part of the stress tensor are both set equal to $\tau
_{r}$, the internal degrees of freedom are eliminated and the approximation $%
\alpha =0$ is made. In order to get a picture of the importance of the
viscoelastic effects in equilibrium, in Fig. 1 we display the equilibrium
spectrum (with $\alpha =0$)\ of both the Maxwell fluid and the Newtonian
fluid. The thermophysical properties, given in Table 1, are those of toluene 
\cite{Toluene}, the temperature has been set to $T_{0}=277$ $K$, we have
taken as wavenumber $k=85200$ $m^{-1}$(which is the value corresponding to a
scattering angle of $10$ $mrad$ and a wavelength of $4880$ $\stackrel{o}{A}$%
) and we have included a very crude estimate of the relaxation time $\tau
_{r}\sim 10^{-5}$ $\sec $ of this liquid obtained from Grad's kinetic
theory expression.

Concerning the nonequilibrium spectrum again with $\alpha =0$, the presence
of the external temperature gradient has a similar effect to the one
observed in the Newtonian fluids for the Brillouin peaks, as illustrated in
Fig. 2. Here we have again used the data corresponding to toluene and taken
the gradient to point upwards with $A=100$ $K/m$, $L=0.01$ $m$ and $T_{0}=277
$ $K$, these latter numbers fulfilling the restrictions involved in our
derivation. A noteworthy feature is that the asymmetry in the Brillouin
peaks, related to the fact that there are more sound waves absorbed with
wavevector $-\mathbf{k}$ than those emitted with wavevector $\mathbf{k}$ due
to the temperature gradient and its corresponding heat flux is much more
pronounced and the peaks become narrower for the viscoelastic fluid. On the
other hand, within the various approximations and simplifying asumptions
that we made, the Mountain peak is not affected by the temperature gradient.
This, which is somewhat surprising and most likely tied to the assumption $%
\alpha =0$, is a trivial consequence of the fact that the nonequilibrium
correction vanishes for $\omega _{c}=0$.

In conclusion, we want to point out that we have derived (to our knowledge
for the first time) the light-scattering spectrum for a Maxwell fluid in a
steady state under the action of an external temperature gradient. Although
many (reasonable) approximations, some of which might be easily disposed of,
were made along the way in order to get analytical results, and in spite of
the fact that the Maxwell fluid is a very simple model of a viscoelastic
fluid, we are persuaded that the present calculation captures the essential
(qualitative) features to be observed in light-scattering experiments in
connection with the elasticity of fluids. Whether this expectation is real
awaits the necessary confrontation with experiment.

\bigskip

\textbf{Acknowledgments.}

This paper is dedicated to Prof. L. S. Garc\'{i}a-Col\'{i}n on the occasion
of his 70th birthday. We want to stress here our indebtness to him for his
many generous contributions to our academic careers. The work of two of us
(M.L.H. and J.A.R.P.) has been partially supported by DGAPA-UNAM\ under
project IN-107798. F. V. acknowledges the partial finantial support by
PROMEP - UAEM. Thanks are also due to Prof. Jan V. Sengers whose work and a
very fruitful conversation a few years ago served as a motivation to carry
out the present research.

Figure Captions.

Fig. 1 Equilibrium light-scattering spectrum as a function of frequency $%
\omega $ for a fluid whose thermophysical data correspond to those of
toluene at $T_{0}=277$ $K$. The scattering wave-number is $k=85200$ $m^{-1}$%
. The dashed line represents the results for the Newtonian fluid and the
continuous line those of the Maxwell fluid.

Fig. 2 Light-scattering spectrum as a function of frequency $\omega $ for a
fluid whose thermophysical data correspond to those of toluene at $T_{0}=277$
$K$ under the action of an external temperature gradient ($A=100$ $K/m$ and $%
L=0.01$ $K$). The scattering wave-number is $k=85200$ $m^{-1}$. The code is
the same as that of Fig. 1.

\bigskip

Table 1

$\eta =0.000553$ $Pa.\sec $

$\eta _{v}=0.00747$ $Pa.\sec $

$\rho _{0}=861.5$ $Kg/m^{3}$

$c=1285$ $m/\sec $

\bigskip

Table Caption.

Table 1. Thermophysical data of toluene as given in Ref. \cite{Toluene}

\begin{figure}[tbp]
\begin{center}
\epsfig{width=10cm,file=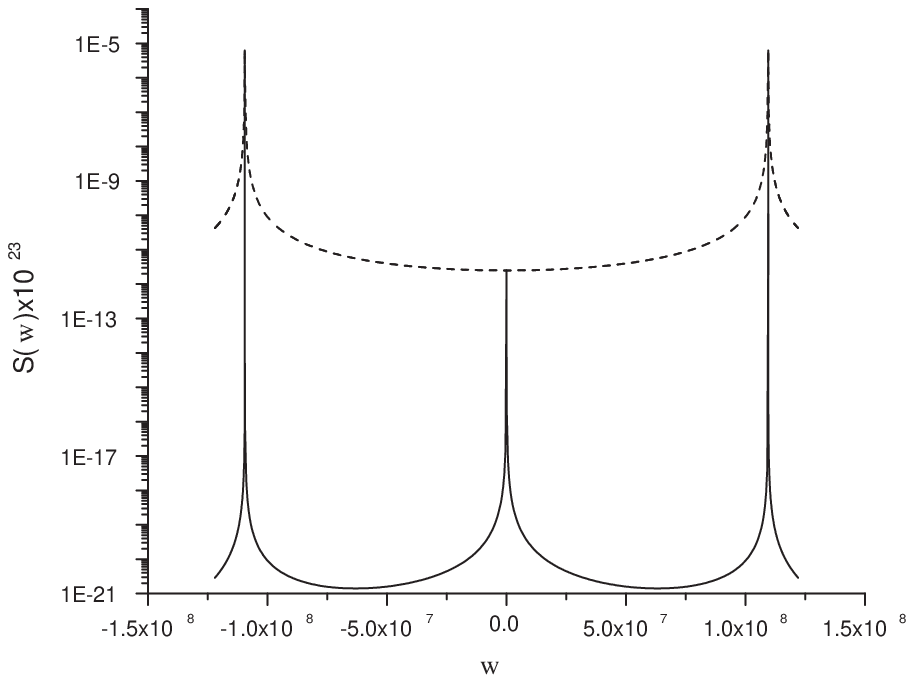}
\end{center}
\end{figure}

\begin{figure}[tbp]
\begin{center}
\epsfig{width=10cm,file=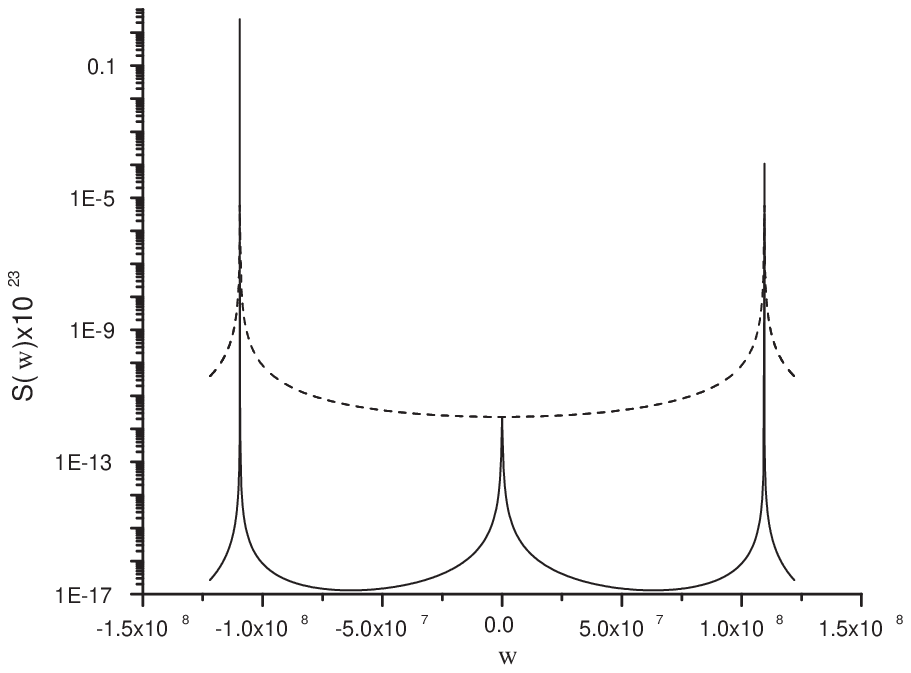}
\end{center}
\end{figure}

\end{document}